\documentclass[num-refs]{nbdt-article}

\usepackage{siunitx}
\usepackage{textgreek}

\papertype{Original Article}
\paperfield{Analysis}

\title{Application of the hierarchical bootstrap to multi-level data in neuroscience}

\author[1]{Varun Saravanan}
\author[2,3]{Gordon J Berman}
\author[2]{Samuel J Sober}

\affil[1]{Neuroscience Graduate Program, Graduate Division of Biological and Biomedical Sciences, Laney Graduate School, Emory University, 30322}
\affil[2]{Department of Biology, Emory University, 30322}
\affil[3]{Department of Physics, Emory University, 30322}

\corraddress{Samuel J Sober, 1510 Clifton Road NE, Room 2006, Atlanta, GA, 30322}
\corremail{samuel.j.sober@emory.edu}

\fundinginfo{The work for this project was funded by NIH NINDS F31 NS100406, NIH NINDS R01 NS084844, NIH NIBIB R01 EB022872, NIH NIMH R01 MH115831-01, NSF 1456912, Research Corporation for Science Advancement no. 25999 and The Simons Foundation.}

\runningauthor{Saravanan et al.}

\begin{document}

\maketitle

\begin{abstract}
A common feature in many neuroscience datasets is the presence of hierarchical data structures, most commonly recording the activity of multiple neurons in multiple animals across multiple trials. Accordingly, the measurements constituting the dataset are not independent, even though the traditional statistical analyses often applied in such cases (e.g., Student’s t-test) treat them as such. The hierarchical bootstrap has been shown to be an effective tool to accurately analyze such data and while it has been used extensively in the statistical literature, its use is not widespread in neuroscience - despite the ubiquity of hierarchical datasets. In this paper, we illustrate the intuitiveness and utility of this approach to analyze hierarchically nested datasets. We use simulated neural data to show that traditional statistical tests can result in a false positive rate of over 45\%, even if the Type-I error rate is set at 5\%. While summarizing data across non-independent points (or lower levels) can potentially fix this problem, this approach greatly reduces the statistical power of the analysis. The hierarchical bootstrap, when applied sequentially over the levels of the hierarchical structure, keeps the Type-I error rate within the intended bound and retains more statistical power than summarizing methods. We conclude by demonstrating the effectiveness of the method in two real-world examples, first analyzing singing data in male Bengalese finches (\textit{Lonchura striata} var. \textit{domestica}) and second quantifying changes in behavior under optogenetic control in flies (\textit{Drosophila melanogaster}).

\end{abstract}

\section{Introduction}
It is commonplace for studies in neuroscience to collect multiple samples from within a category (e.g., multiple neurons from one animal) to boost sample sizes. A recent survey found that of 314 papers published in prominent journals covering neuroscience research over an 18 month period in 2013-14, roughly 53\% of those studies had nested datasets featuring hierarchical data~\cite{aarts2014solution}. When data are collected in this manner, the resulting data are not independent. Commonly deployed statistical tests like the Student’s t-test and ANOVA, however, treat all data points as independent. This assumption results in an underestimation of uncertainty in the dataset and a corresponding underestimation of the p-value~\cite{hahs2005primer,arceneaux2009modeling,musca2011data}. Pseudoreplication~\cite{hurlbert1984pseudoreplication,lazic2010problem}, where variance within a group and variance between groups is not appropriately accounted for, is often the cause of such biases. For example, consider a hypothetical example in which one measures changes in dendritic spine size during learning. Since one can typically only measure from a few animals each in different treatment conditions, researchers usually increase sample sizes by measuring multiple spines from each neuron and by measuring multiple neurons within an animal. The hierarchical nature of such datasets can result in different samples being statistically dependent on each other: spines measured from the same neuron are likely more similar than spines measured across different neurons, even more so than spines measured from different animals within the same treatment condition. 

Linear Mixed Models (LMMs) can be used to account for the variance across different levels~\cite{aarts2014solution,aarts2015multilevel} and have recently been used to do so in several studies~\cite{arlet2015grooming,liang2015mapping,machado2015quantitative,pleil2016effects}. However, LMMs assume that all hierarchical structure present is linear, which is often not the case. Additionally, the parameters returned by LMM fits may exhibit bias and be unreliable when the number of clusters is small, as is also often the case in neuroscience datasets~\cite{maas2005sufficient,gehlbach2016creating,huang2018using}. Finally LMMs provide a great deal of flexibility in the various implementation choices a user can make to build models appropriate for their datasets. However, recent findings have suggested that LMMs can be highly sensitive to user choices, particularly for random effects, which in turn can impact the convergence and interpretation of results from LMMs~\cite{seedorff2019maybe}.

The hierarchical bootstrap~\cite{efron1982jackknife,efron1992bootstrap,efron1994introduction,carpenter2003novel} is a statistical method that has been applied successfully to a wide variety of clustered datasets, including census and polling results, educational and psychological data, and phylogenetic tree information~\cite{efron1996bootstrap,harden2011bootstrap,huang2018using}. Unlike LMMs, the hierarchical bootstrap is relatively agnostic to the underlying structure present in the data and has consistently performed better at quantifying uncertainty and identifying signal than traditional statistics~\cite{field2007bootstrapping,harden2011bootstrap,thai2013comparison}, though some concerns have been raised that the bootstrap may be excessively conservative in a limited subset of cases~\cite{hillis1993empirical,adams1997resampling}. However, the use of the hierarchical bootstrap in neuroscience has been limited, even though its application is increasingly warranted.

This paper is divided into two parts. In the first, we simulate a typical dataset studied in neuroscience and use it to illustrate how the Type-I error is inflated in hierarchical datasets when applying traditional statistical methods but can be averted using the hierarchical bootstrap. In the second, we demonstrate the use of the hierarchical bootstrap in two real-world examples using singing data from songbirds~\cite{hoffmann2014vocal} and optogenetic control of behavior in flies~\cite{cande2018optogenetic}. In both cases, the data have a strong hierarchical structure and our analyses highlight the need to use appropriate statistical tests when analyzing hierarchical datasets in neuroscience.

\section{Materials and Methods}
The simulations for this paper were run in the Jupyter Notebooks environment using Python (version 3.7.2) and importing the following libraries: NumPy (version 1.15.4), SciPy (version 1.1.0), Matplotlib (version 3.0.2), StatsModels (version 0.11.0) and Pandas (version 0.23.4). Reanalysis of data from Hoffmann and Sober (2014) was performed using MATLAB (version 2017a). The codes for both simulation and the data analysis are available on Github here: \href{https://github.com/soberlab/Hierarchical-Bootstrap-Paper}{https://github.com/soberlab/Hierarchical-Bootstrap-Paper}.

\begin{figure}[!ht]
\centering
\includegraphics[width=10.5cm]{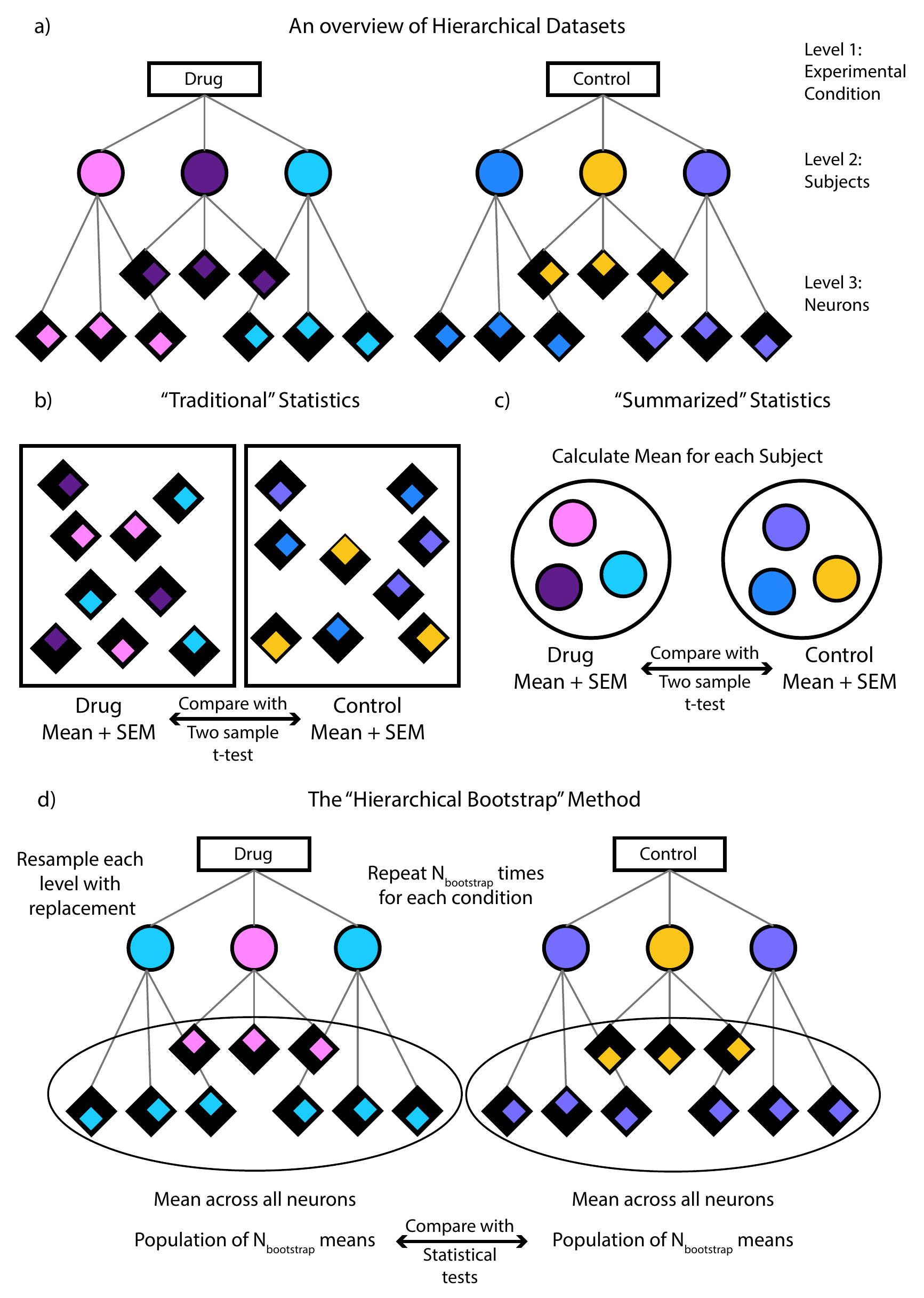}
\caption{a) An example of a hierarchical dataset. Here the dataset is divided into 3 levels, with the first level containing the experimental groups to be compared, the second containing the individual subjects and the third containing the individual neurons per subject. Each subject is color coded and the neurons per subject are distinguished by the position of the colored diamond. b) In “Traditional” statistics, the means for each group is computed across all the neurons and are then compared using a two sample t-test. c) In “Summarized” statistics, the mean for each subject is computed first. These means are then used to compute an overall mean for each group and the groups are compared using a two-sample t-test. d) In the “Hierarchical Bootstrap” method, we create new datasets N\textsubscript{bootstrap} times by resampling with replacement first at the level of subjects followed by neurons within a subject. We then compute the mean across all neurons every time we perform resampling. The final statistic is computed on this population of resampled means (see Methods for details).}\label{fig:HierDatSchem}
\end{figure}

\subsection{Traditional vs Summarized vs Bootstrap}
Throughout this paper, we compare 4 statistical methods that we refer to by shorthand as “Traditional”, “Summarized”, “Bootstrap” and "LMM" respectively. Here, when we refer to the “Bootstrap” method, we mean a hierarchical non-parametric bootstrap procedure. We will detail what the first three of those terms mean here (see Fig.~\ref{fig:HierDatSchem} for schematics of each), and we will describe LMMs in detail in a subsequent section. For the sake of clarity, let us consider a fictitious example. Suppose our dataset involves recording the neural activity of neurons in the amygdala when subjects were exposed to an aversive auditory cue either in the presence or absence of a drug of interest believed to reduce anxiety. For each subject, neural activity was recorded from one hundred neurons during exposure to the auditory cue and the experiment was repeated with several hundreds of subjects in both the presence and absence of the drug (see Fig.~\ref{fig:HierDatSchem}a). Note that while a more typical example would involve only a small number of subjects, the differences between methods are easier to comprehend in an example with a large number of subjects (we demonstrate differences in a small number of subjects in Fig.~\ref{fig:Power}). We could add a layer of complexity by considering that the experiment was repeated for several trials for each neuron recording but for the sake of simplicity, let us assume that all the data were collected from a single trial per neuron. In the “Traditional” method, every data point (e.g., the firing rate of every neuron to every instance of the auditory cue) is treated as independent, regardless of the hierarchical structure present in the dataset (see Fig.~\ref{fig:HierDatSchem}b). All the data points are used to calculate the mean and  the uncertainty in the estimate of the mean, namely the standard error of the mean (SEM) and a Student’s t-test is used to ascertain statistically significant differences between the mean firing rate of the neurons in the presence versus absence of the drug of interest. This is fairly common since, unlike the previous fictitious example, experiments are often repeated only in a small sample of subjects. The “Summarized” method, on the other hand, acknowledges the possibility that different neurons within the same subject may be more similar to each other than neurons across subjects. As a result, the mean firing rate for each subject is calculated first and the mean of the group is calculated as the mean of the population of mean firing rates for each subject in the group and the SEM is computed from this population of means (see Fig.~\ref{fig:HierDatSchem}c). Note that the mean for each group in this case is equal to that in the “Traditional” case if and only if the number of neurons recorded for each subject within a group is represented equally. A Student’s t-test is thus applied to the population of mean firing rates between the two groups. An additional complication that we circumvent in our toy example by not considering trials per neuron is the decision as to which level one must summarize the data. In the case of multiple trials per neuron, one may summarize either at the level of individual neurons or individual subjects. While summarizing at the level of subjects is the most appropriate way to avoid non-independence between data points, it can seriously reduce sample size and therefore power especially if the number of subjects is small. In the “Bootstrap” method, we perform the hierarchical bootstrap on the two highest level groups (drug versus control group in Fig~\ref{fig:HierDatSchem}d) to compute posterior distributions of the range of means possible from each group, as follows. First, we sample with replacement (i.e., we sample from the current distribution in such a way that replications of previously drawn samples are allowed) from the subjects in the group. Then, for the subjects selected, we then sample with replacement from the individual neurons for the number of neurons that were recorded. We then compute the mean firing rate across the group for that resampled population and repeat the entire process N\textsubscript{bootstrap} times (N\textsubscript{bootstrap}=10\textsuperscript{4} for all instances in this paper unless otherwise noted). The mean for each group in this case is computed the same way as is done in the “Traditional” method. The 67\% confidence interval (or equivalently, the standard deviation) of the population of N\textsubscript{bootstrap} means gives an accurate estimate of the uncertainty in the mean. Note that the mean is a special case where this uncertainty is more commonly referred to as the Standard Error of the Mean or SEM. We can then compute the probability of one group being different from the other using the population of resampled means obtained above for each group (see Hypothesis testing with Bootstrap below for complete details). 

\subsection{Hypothesis testing using Bootstrap}
We described above how bootstrap samples can be used to compute the uncertainty in measuring the mean of a population. However, the bootstrap can be used more broadly to measure the uncertainty in any metric of interest as long as it obeys the law of large numbers and scales linearly in the probability space. In addition, the bootstrap is used to compute posterior distributions of the range of values possible for the metric of interest from the data of limited sample size. As a result, the distribution of bootstrap samples can be used to compute probabilities that the data supports particular hypotheses of interest directly as opposed to p-values (see below for a detailed description). We will describe below how this can be done with an example. Note that while this is not the only way of hypothesis testing using bootstrapping, we found this to be a particularly simple and effective way of doing so.

As will be performed several times in this paper, suppose we wish to evaluate the support for the hypothesis that the mean of particular sample was significantly different from a fixed constant - zero for our example. In order to do so, we would compute the proportion of means in our population of bootstrapped means of the sample that were greater than or equal to zero. If we set the acceptable false positive (Type-I error) rate to \textalpha ~(\textalpha = 0.05 throughout this paper), then if the computed proportion was greater than 1 – \textalpha/2 (or p>0.975 for \textalpha = 0.05), we would conclude that the sample of interest had a mean significantly greater than zero. Alternatively, if the computed proportion was less than \textalpha/2 (or p<0.025 for \textalpha = 0.05) then we would conclude that the sample of interest had a mean significantly less than zero. Any proportion $\alpha/2 \leq p \leq (1 - \alpha/2)$ would indicate a relative lack of support for the hypothesis that the mean of the sample of interest is different from zero. In the case of comparing the population of bootstrapped means to several constants (multiple comparisons), we use the Bonferroni correction to adjust the threshold for significance accordingly. We would also like to make a distinction between the probabilities we referred to above and the p-values typically associated with statistical tests. p-values refer to the probability of obtaining a result as extreme or more extreme than those obtained under the assumption that the null hypothesis is true. As such, they do not provide a direct measure of support for the hypothesis one truly wishes to test. The ‘p’ referred to in the bootstrapping procedure above, however, directly provides a probability of the tested hypothesis being true. For the rest of this paper, in order to distinguish the direct probabilities obtained using the bootstrapping procedure from p-values reported from traditional statistical tests, we will use ‘p\textsubscript{boot}’ to refer to bootstrap probabilities and ‘p’ to refer to p-values from other tests.

The procedure described above can also be used to compare the means of two different groups using their respective samples (although this variant of the analysis is not performed in this manuscript). In this case, we would compute a joint probability distribution of the two samples with each sample forming the two axes of a 2-D plot. The null hypothesis would thus be a symmetric Gaussian with the line y = x as a diameter. Therefore, to test if the two groups are different, one would compute the total density of the joint probability distribution on one side of the unity line. If the volume computed is greater than 1 – \textalpha/2 then the first group is significantly greater than or equal to the second while if the volume computed is less than \textalpha/2, the second group is significantly greater than or equal to the first with all other volumes indicating no significant differences between the groups. We can also extend this formulation to comparisons between multiple groups by performing pairwise-comparisons between the groups and adjusting the threshold for significance accordingly (using Bonferroni corrections~\cite{weisstein2004bonferroni}, for example).

\subsection{Choosing the number of times to resample using Bootstrap}
As we mentioned in the Introduction, the bootstrap is relatively agnostic to the structure of the data. However, this statement is not to say that there are no assumptions in its use. As mentioned previously, the bootstrap cannot be used on a metric that does not obey the law of large numbers (i.e., the central limit theorem) or is not linear in the probability space (e.g., depends on the product or logarithm of probabilities like entropy for example). Additionally, how one chooses to resample their data defines assumptions that must match the true data distribution in order to provide accurate results. A common procedure is to resample from a population equal to the number of samples drawn in the original dataset. While this may seem obvious at higher levels of hierarchical datasets, it is not so clear at the lower levels. To take the example of songbird vocal data, one bird may simply have sung far fewer times than the rest of the birds in the experimental group. Should that bird’s data always be less represented in all resampled datasets? What if the imbalance is in the number of repetitions of individual syllables within a bird? In each case, the way to resample may and should change based on the hypothesis being tested. The crucial rule to follow when choosing resampling parameters is to come as close as possible to replicating what one might expect if they had physically repeated that experiment several hundreds of times instead of resampling using bootstrapping. For our own work, we have previously given equal weight to each syllable independent of its frequency of occurrence in the actual dataset~\cite{saravanan2019dopamine} while in the experimental results analyzed here, each syllable is weighted by its frequency of occurrence due to the differences in the nature of experiments and questions being analyzed.

\subsection{Linear Mixed Models (LMMs)}
Linear Mixed Models have been suggested and successfully used as a powerful technique to account for various datasets that have a non-independent structure including hierarchical datasets and time series data~\cite{snijders2011multilevel,hox2017multilevel}. We have therefore included results from the use of LMMs in our simulations and reanalysis as we believe it will provide useful context as to when one may choose to use one technique over the other. LMMs are extensions of linear regressions in that they consist of a “fixed effect” component and a “random effect” component. The “fixed effect” component is the conventional linear regression part and typically is what one uses to quantify statistical differences between groups. The “random effects” on the other hand can vary across units (neurons, subjects, etc.) in the dataset allowing one to build in the hierarchical structure of the data into their model. In our simulations, we used the treatment condition as the fixed effect and subject identity as the random effect. The proportion of significant results was obtained by extracting the t-statistic and corresponding p-value for the fixed effect coefficient. The SEM estimate was drawn from the standard error in the intercept for the model. We describe the LMMs used in our experimental work in greater detail in the corresponding section under Results.

\subsection{Design Effect (DEFF)}
In scientific studies, the effect size is an important metric that helps define the effectiveness of the treatment and the number of subjects required to adequately power the study. It is therefore important to quantify changes in the effect size due to the presence of a hierarchical structure in the dataset. When one analyzes data from hierarchical datasets, the unique information provided by each additional data point at the lowest level of the hierarchy depends on the average number of samples in the cluster (neurons within a subject form a cluster in our toy example) and the relative variance within and between clusters. This relationship was mathematically quantified using the Intra-cluster correlation (ICC). ICC is a useful metric that provides a quantitative measure of how similar data points are to each other within an individual cluster in a hierarchical dataset~\cite{walsh1947concerning,kish1965survey}. While there are some differences in how it is calculated, in general it is defined as the following ratio:
\begin{equation}
ICC \quad or \quad \rho = \frac{s_{between}^2}{s_{between}^2 + s_{within}^2}    
\end{equation}
Here $s_{between}^2$ represents the variance across the means of individual clusters and $s_{within}^2$ represents the variance within clusters. Thus, the ICC is a metric that varies from zero to one, where a measure of zero represents no clustering of data and every data point being independent and a measure of one represents a perfect reproduction of samples within clusters (i.e., all points within a cluster are exactly the same). Kish further formalized the relationship between ICC and the adjusted effect size that was termed the “Design Effect” or DEFF with a corresponding correction to be applied to the standard error of the mean computed from the dataset termed DEFT, defined as the square root of DEFF~\cite{kish1965survey,mccoach2010dealing}. Formally, DEFF is defined as:
\begin{equation}
    DEFF = \frac{\sigma^2(data)}{\sigma^2(data\ if\ independent)} = 1 + \rho*(\bar{n}_j - 1)
\end{equation}
Where \=n\textsubscript{j} represents the average sample size within each cluster and \textrho~ is the ICC. Hence, as the number of samples within a cluster increases, the DEFF increases, resulting in a need for a larger correction (increase) to the standard errors. Conversely, as the number of samples within clusters increase, the standard error of the mean is underestimated potentially resulting in underestimation of the p-values and inflation of the Type-I error rate.

\section{Results}
Our results are organized into two sub-sections: Simulations and Examples. In the Simulations sub-section, we show results from simulations that illustrate the utility of the hierarchical bootstrap and in the Examples sub-section, we highlight the differences in results when analyzing two example datasets with and without the hierarchical bootstrap and also compare to LMMs. Throughout the results section, we will compare statistical tests we refer to by shorthand as “Traditional”, “Summarized”, “Bootstrap” and "LMMs" respectively (see \textit{Traditional vs Summarized vs Bootstrap} and \textit{Linear Mixed Models (LMMs)} in \textit{Materials and Methods} and Figure~\ref{fig:HierDatSchem} for a detailed description). Also note that whenever we refer to the “Bootstrap” in this paper, we mean a hierarchical non-parametric bootstrap unless otherwise specified.

\subsection{Simulations}
We used simulations of neuronal firing in order to highlight the key characteristics of the hierarchical bootstrap as applied to nested data in neuroscience and the differences between the bootstrap and other, more commonly used statistical tests. Since it has been a concern raised previously~\cite{hillis1993empirical,adams1997resampling}, we were interested in whether the bootstrap displayed a conservative bias for independent (i.e., all data points are as dissimilar to each other as to every other data point) and non-independent (i.e., some data points are more similar to each other than others) datasets, as well as in quantifying the bootstrap’s statistical power compared to other techniques. While the results of these simulations may be predicted by other mathematical results previously published~\cite{davison1997bootstrap,carpenter2003novel}, we found them instructive to depict explicitly.

\subsubsection{The hierarchical bootstrap does not have a conservative bias in a hierarchical dataset}
\begin{figure}[!ht]
\centering
\includegraphics[width=14cm]{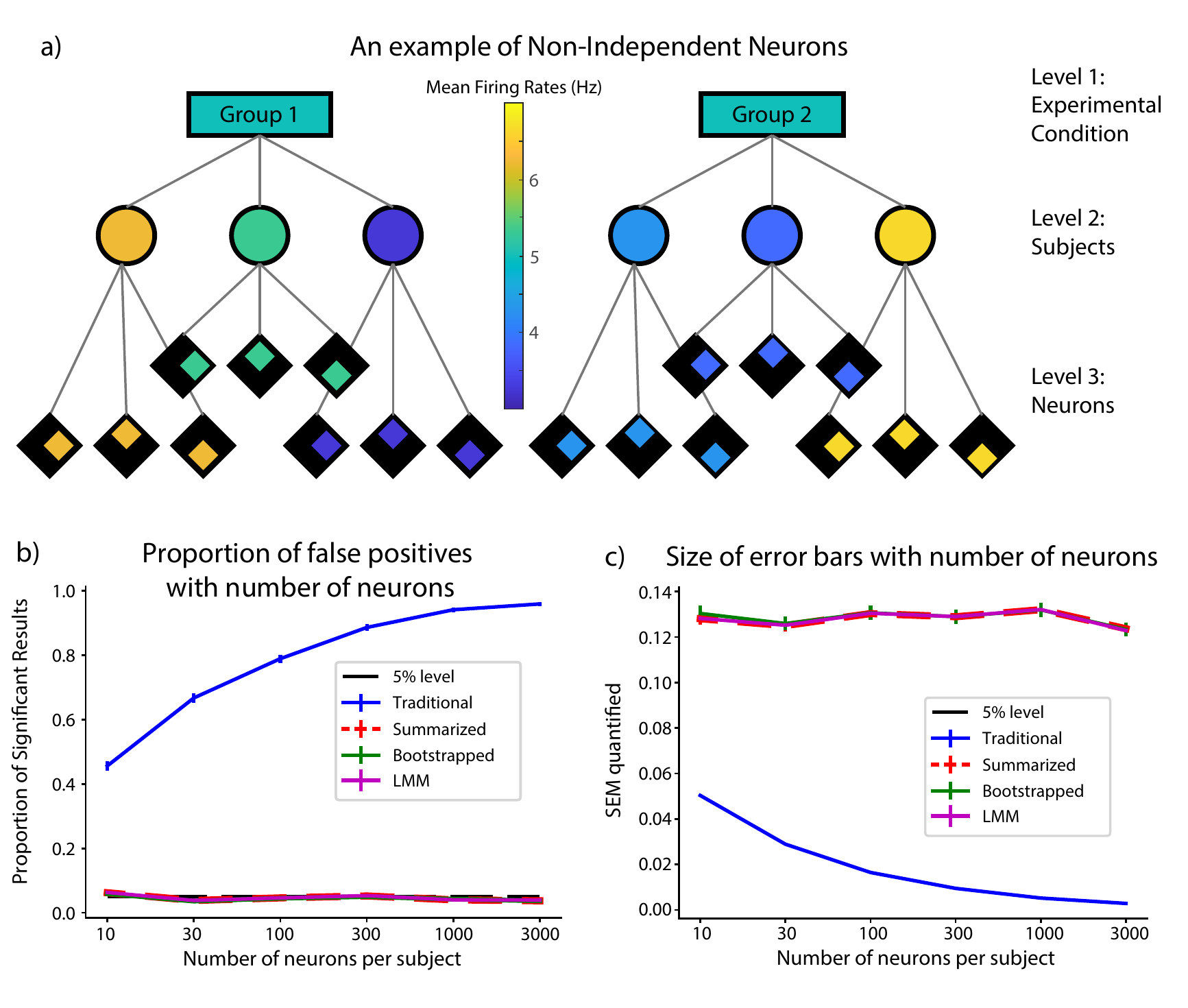}
\caption{False positive rate and size of error bars quantified for the simulation in which there was no difference between the groups but the points were not independent. a) A graphical representation of the experimental condition. The firing rate is not different between the two groups but each subject has a constant offset away from the mean for the group that is consistent across all their neurons. b) The proportion of significant results using each statistical method as a function of the number of neurons per subject. As expected, the false positive rate for the traditional method rises with increasing number of samples within each subject. The Summarized, Bootstrap and LMM methods on the other hand have almost identical false positive rates not significantly different from the theoretical 5\% in all cases. c) The size of SEMs (uncertainty in computing the mean firing rate for each group) computed for all 4 methods as a function of the number of neurons per subject. While those for the summarized, bootstrap and LMM stay roughly the same, those for the traditional method reduces with increasing number of trials. Note that for both traces, since the Bootstrap, Summarized and LMM almost perfectly overlap, the red trace has been thickened and dashed for visualization.}\label{fig:NullDep}
\end{figure}

If the bootstrap had a strong conservative bias regardless of the nature of the data (hierarchical or independent), it may not be the right metric with which to address the problem of statistical analysis in hierarchical datasets. For our first simulation, we wished to evaluate whether the bootstrap displayed a conservative bias when processing a typical hierarchical dataset one might encounter in neuroscience research. Specifically, we simulated a condition in which we recorded the activity of 100 neurons each from 1000 subjects. The subjects would then be split randomly into two groups of 500 each and the mean firing rate across groups would be compared. Each neuron had a mean firing rate of 5Hz, simulated using a Poisson process. However, in order to introduce hierarchical structure into the dataset, each subject’s neuronal firing rate was offset by a constant drawn from Gaussian noise of width 3Hz. This constant was the same for all 100 neurons simulated for each subject but varied between subjects. Since subjects were split randomly into two groups, there should be no statistical difference between the groups. Therefore, any significant differences we see would be false positives and we would expect to see them at a rate of \textalpha (here set to 0.05). This has been depicted graphically in Figure~\ref{fig:NullDep}a where the two groups have identical mean firing rates but the individual subjects have an offset that is constant across all their neurons. 

We also varied the number of neurons per subject to study its effect on the false positive rate. We simulated the experiment 1000 times for each value of number of neurons and computed the false positive rate from each. We used bootstrapping on the obtained results to estimate error bars and to test for significant differences away from 0.05, the expected false positive rate. Given the relationship between the number of points within a cluster to the Design Effect (DEFF; see \textit{Design Effect} in Materials and Methods), we would expect the false positive rate to increase with the number of neurons per subject in the traditional analysis but not using the other methods~\cite{snijders2011multilevel,snijders1993standard,aarts2014solution}. 

As shown in Figure~\ref{fig:NullDep}b, the false positive rate for the traditional method does increase with the number of neurons per subject rising from around 46\% for 10 trials to almost 96\% in the case of 3000 trials per neuron (probability of resampled proportions being greater than or equal to 0.05 was p\textsubscript{boot} > 0.9999 in all cases; limit due to resampling 10\textsuperscript{4} times). On the other hand, the summarized, bootstrap and LMM methods stayed remarkably similar in value and were not significantly different from 0.05, the expected false positive rate, in all cases (adjusting for threshold of significance with Bonferroni corrections for 3 comparisons).

We also computed the estimate for the SEM, i.e., the uncertainty in computing the mean firing rate for each group, using all 4 cases for each number of neurons simulated, and the result is shown in Figure~\ref{fig:NullDep}c. As shown, the SEM estimate remains fairly constant for the summarized, bootstrap and LMM methods but decreases with an increase in the number of trials per neuron in the traditional case. Furthermore, the SEM estimate for the traditional case starts out much lower than the other cases suggesting that the increased false positive rate is at least partially due to the underestimation of error in the traditional case.

\subsubsection{The hierarchical bootstrap is more conservative than Traditional and Summarized methods for independent data points but LMM is not}
In the previous simulation, we reported that the hierarchical bootstrap does not have a conservative bias for hierarchical datasets. However, it has been claimed that the bootstrap has a conservative bias~\cite{hillis1993empirical,adams1997resampling}, resulting in larger error bars than strictly necessary for the chosen threshold of Type-I error \textalpha~(here set to 0.05). It has also been argued that this is not a bug or bias in the algorithm, but rather a more generic property of hypothesis testing by resampling~\cite{felsenstein1993there,efron1996bootstrap}, and newer algorithms have claimed to reduce bias further~\cite{shimodaira2002approximately,shimodaira2004approximately}. Here we tested the conservative bias of the hierarchical bootstrap in a similar situation as the first simulation above but where all the data points were independent. Given that we set \textalpha~to 0.05, we would expect a 5\% false positive rate if there were no bias in the algorithm.

\begin{figure}[!ht]
    \centering
    \includegraphics[width=14cm]{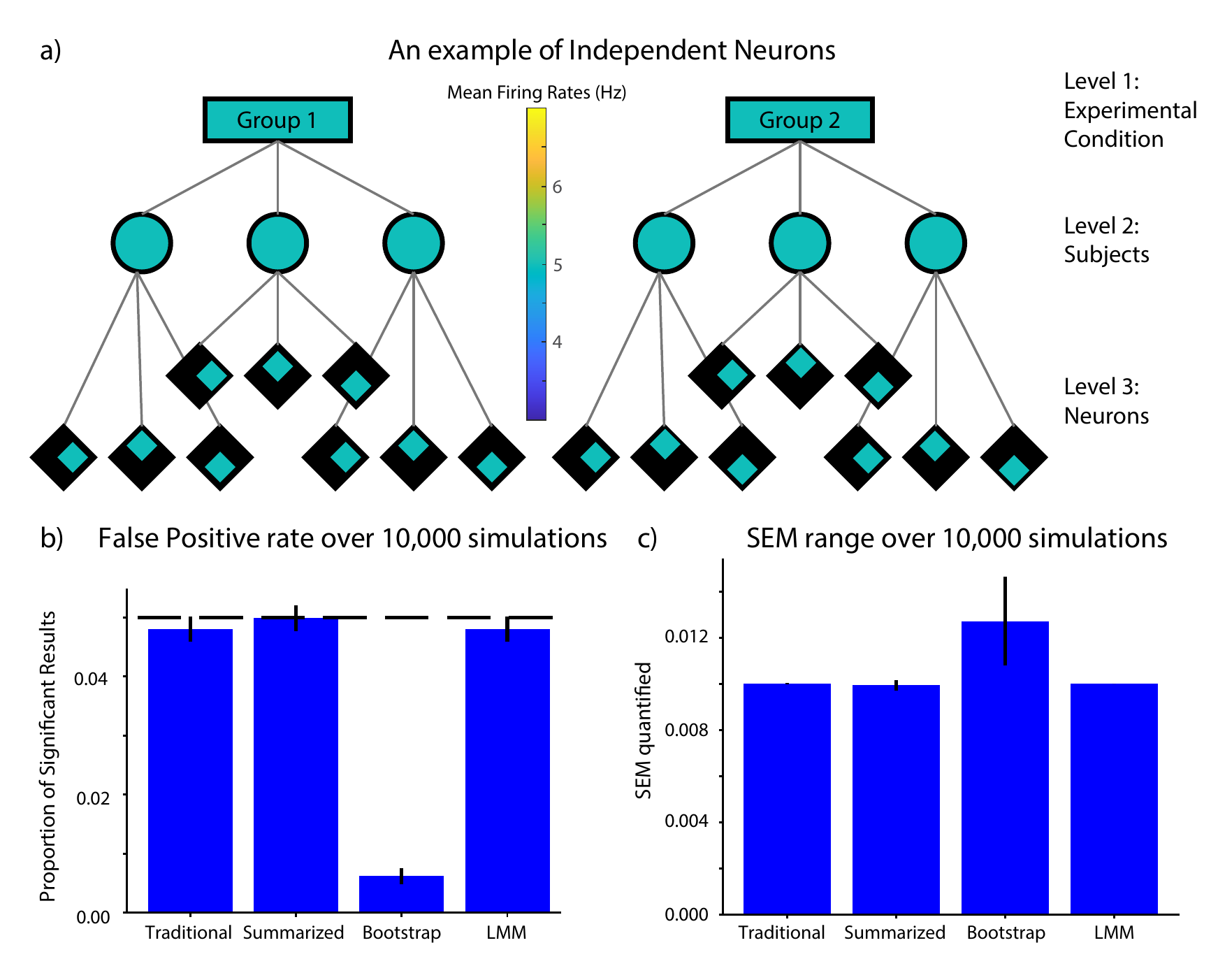}
    \caption{Results from the simulation in which all data points were independent. a) A graphical representation of the experimental condition. As shown by the shades of group, subject and neuron means all being the same, the points are all independent in spite of an apparent hierarchical structure. b) Proportion of significant results when comparing the 2 groups with each statistical method at \textalpha of 0.05. As expected, the Traditional, Summarized and LMM methods give roughly 5\% false positive results (black dashed line). However, the bootstrap gives a much smaller proportion of significant results suggesting a conservative bias. c) The size of SEMs computed using each of the methods. The bootstrap does give an error bar roughly 1.4 times that of the other metrics.}
    \label{fig:NullIndep}
\end{figure}

As before we simulated a situation in which we recorded the activity of 100 neurons each over 1000 subjects. In order to make each neuron independent though, we removed the Gaussian noise term for each subject. The neurons were thus simulated using only a Poisson random number generator with an average firing rate of 5Hz (each trial was thought to be 1 second of activity). This has been depicted graphically in Figure~\ref{fig:NullIndep}a where the group means, subject means and trials all are set to the same firing rate of 5Hz. As is evident from the figure, even though there is a hierarchical structure present, the points can be rearranged in any fashion without affecting the results. In other words, the points are all independent. Since the data points are independent, we would not expect differences between the Traditional and Summarized methods. We then split these 1000 subjects into two groups of 500 each randomly and computed the mean firing rate for each group. We then tested whether the means were significantly different from each other using the Traditional, Summarized, Bootstrap and LMM methods. We repeated this analysis 10000 times and plotted the proportion of simulations that resulted in significant differences for each of the methods in Figure~\ref{fig:NullIndep}b. The error bars were computed by bootstrapping the results obtained from the simulation runs. As shown in the figure, the Traditional, Summarized and LMM methods resulted in a proportion of significant results close to and not significantly different from 5\% as expected (Traditional – 4.80 $\pm $ 0.21 \%; probability of proportion of significant results being greater than or equal to 0.05 was p\textsubscript{boot} = 0.18; Summarized – 4.99 $\pm $ 0.22 \%; probability of proportion of significant results being greater than or equal to 0.05 was p\textsubscript{boot} = 0.48; LMM – 4.80 $\pm $ 0.21 \%; probability of proportion of significant results being greater than or equal to 0.05 was p\textsubscript{boot} = 0.18). By contrast, when using the bootstrap method, the proportion of significant results was significantly lower than the expected 5\% at 0.66 $\pm $ 0.08 \% (probability of proportion of significant results being greater than or equal to 0.05 was p\textsubscript{boot} < 10\textsuperscript{-4}; limit due to resampling 10\textsuperscript{4} times). This was a marked departure from Figure~\ref{fig:NullDep}b where we saw that the bootstrap had no significant conservative bias for hierarchical datasets.

We also computed the standard error of the mean (SEM) in each case and reported the results in Figure~\ref{fig:NullIndep}c. As shown, the error bars for the Traditional, Summarized and LMM methods are almost identical at 1.002 $\pm $ 0.002 * 10\textsuperscript{-2} for Traditional, 0.994 $\pm $ 0.023 * 10\textsuperscript{-2} for Summarized and 1.001638 $\pm $ 0.000008 * 10\textsuperscript{-2} for LMM respectively. The error bars computed using the Bootstrap method are roughly 1.4 times larger at 1.407 $\pm $ 0.035 * 10\textsuperscript{-2}. Since the effect size is inversely proportional to the uncertainty in the dataset~\cite{coe2002s}, which is captured here by the error bars, we conclude that the larger error bars do partially account for the drop in proportion of significant results observed and that the bootstrap seems to have a conservative bias for independent datasets.

\subsubsection{The Bootstrap and LMM balance intended Type-I error rate and statistical power better than the Traditional or Summarized methods at low sample sizes}
\begin{figure}[!ht]
    \centering
    \includegraphics[width=14cm]{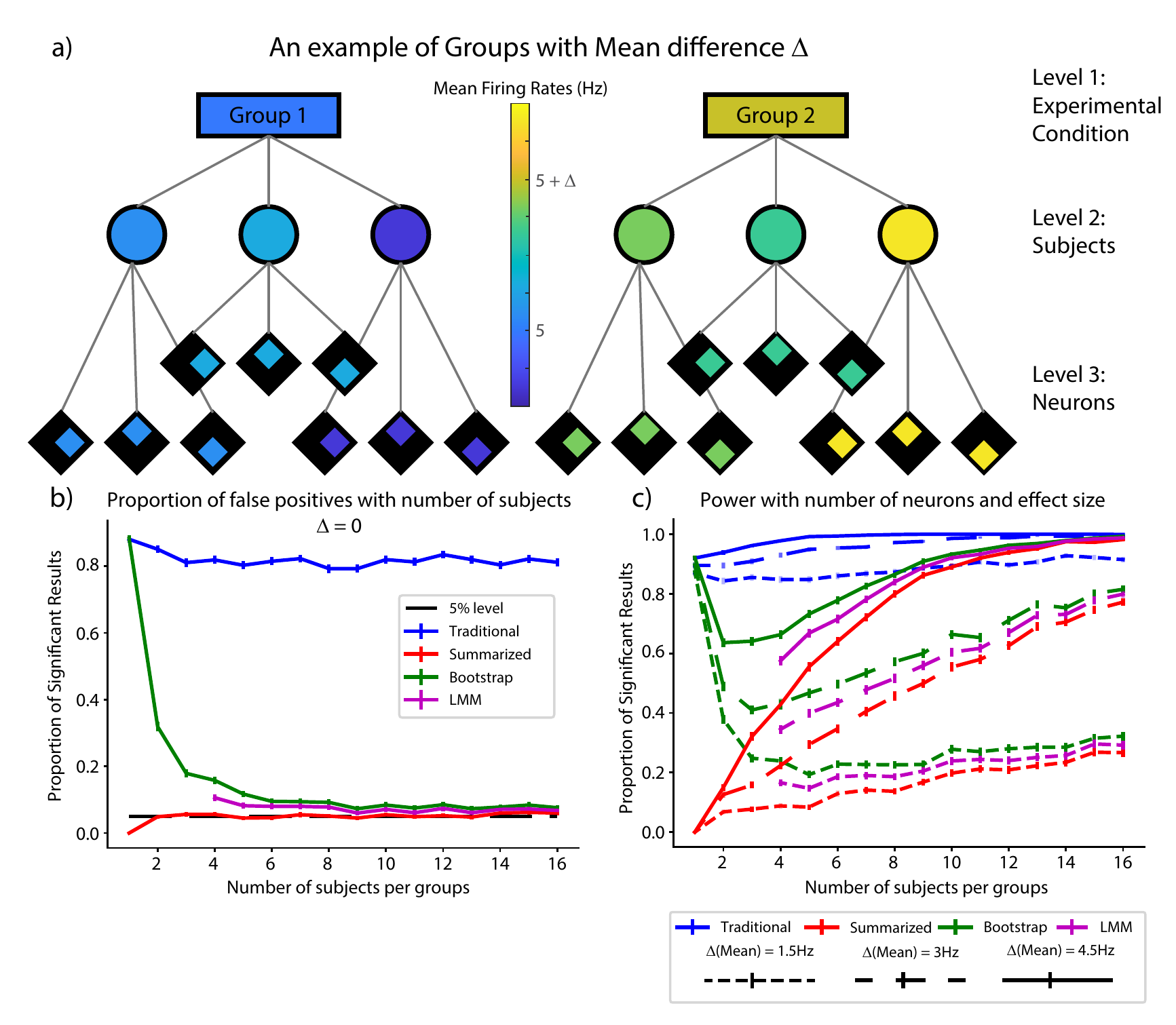}
    \caption{Change of power with number of neurons and effect size. a) A graphical representation of the experimental conditions. As shown, there is now a difference in the mean between groups with individual subjects also displaying variations about the mean. b) The false positive rate when there was no difference between the mean firing rates for the two groups. c) The proportion of significant results or power when the difference in mean firing rates (\textDelta mean) between the two groups was 1.5Hz (short dashes), 3Hz (long dashes) and 4.5 Hz (solid line) respectively.}
    \label{fig:Power}
\end{figure}

As we saw in the previous section, the summarized, bootstrap and LMM methods bind the Type-I error rate at the intended 5\% and the estimate of the SEM is roughly the same for all three methods (Fig.~\ref{fig:NullDep}). What then is the advantage of the bootstrap or LMM over simply using the summarized method? The answer lies in the fact that the summarized methods result in a loss of statistical power (the ability to detect a statistical difference when one truly exists), particularly for low sample sizes of the upper hierarchical levels and for small effect sizes (a situation commonly found in neuroscience datasets). We used simulations to calculate the power for each of the four methods and the results are shown in Figure~\ref{fig:Power}.

Since power depends on the effect size and the number of samples, we chose to examine the change in power with respect to the number of subjects per group (N) and the effect size for these simulations (\textDelta mean). In order to do so, we varied N between 1 and 16, keeping the number of neurons per subject constant at 100 each. As shown in Figure~\ref{fig:Power}a, we kept the mean firing rates of neurons of one group of subjects at 5Hz as before and varied the mean firing rate of the other group by \textDelta mean, adding an additional 3Hz random Gaussian noise term to each subject in both groups. As before, this term is constant for all neurons within a subject and varies between subjects. Since the previous simulations did not estimate the false positive rate when the number of subjects was as low, we first kept the mean firing rate for both groups of subjects equal at 5Hz, simulating 1000 times for each value for the number of subjects per group. The result is shown in Figure~\ref{fig:Power}b. As shown, the false positive rate for the traditional method stays around or above 80\% (blue trace in Fig.~\ref{fig:Power}b), while that for the summarized method hugs the expected 5\% line (red trace in Fig.~\ref{fig:Power}b), except for the special case of one neuron per group, where you can never achieve significance since you are comparing only two points. The behavior of the bootstrap calculation highlights the fundamental characteristic of the bootstrap and is therefore worth exploring in detail (green trace in Fig.~\ref{fig:Power}b). The essence of the bootstrap is to provide a reliable range for your metric under the assumption that the limited dataset you have captures the essential dynamics of the underlying population. When there is only one subject, the bootstrap assumes that neural level data is the true distribution and therefore has a false positive rate equal to that of the traditional method. As the number of subjects increases, one gets a better sampling of the true underlying distribution, and correspondingly, the bootstrap tends towards a 5\% error rate with increasing number of subjects, as the weight of data points shifts from individual neurons to neurons across subjects with increasing number of subjects. Therefore, if the data collected do not accurately represent the dynamics of the underlying distribution, the bootstrap cannot provide accurate estimates of population metrics. The LMMs also have a false positive rate that is higher than the expected 5\% rate for very small N which also reduces with increasing N (magenta trace, Fig.~\ref{fig:Power}b). However, it is consistently less than that of the bootstrap. Note that for LMMs, the algorithm does not converge reliably for very small group numbers and so we could only plot results for N $\geq$ 4. In spite of this, we verified that the false positive rate for LMMs at N=1 is 16 $\pm $ 1\%, well below that of the bootstrap.

We then computed the power for the four methods as a function of the number of subjects per group and the difference in mean firing rate between the groups. Accordingly, we repeated the simulations described above changing the mean firing rate of one of the groups to 6.5Hz, 8Hz and 9.5Hz (short dashed, long dashed and solid traces in Fig.~\ref{fig:Power}c respectively). Since there is an actual difference between the groups in this case, the ideal plot will have a very high proportion of significant results barring adjustments for extremely low sample sizes. As shown, the traditional method has the most power (blue traces in Fig.~\ref{fig:Power}c), but as was seen in Figure~\ref{fig:Power}b, also has an unacceptably high false positive rate for this type of data. The summarized method has the lowest power among the three methods, but does catch up for large effect sizes and with increasing group sizes (red traces in Fig.~\ref{fig:Power}c). Both the bootstrap and the LMM are between the two extremes and have more power than the summarized metric particularly for small effect sizes and small group sizes (green and magenta traces in Fig.~\ref{fig:Power}c respectively). As a result, we see that the bootstrap and the LMM help retain statistical power while also being sensitive to the Type-I error rate with the bootstrap being marginally better. However, as was mentioned when discussing Figure~\ref{fig:Power}b, the bootstrap can weight trials within levels more heavily than one would expect if the number of samples in the upper levels is very low, more so than LMMs, and one must therefore be mindful when dealing with very low sample sizes that their data collected may not represent the true distribution in the population. Overall, our simulations suggest that both the bootstrap and the LMM perform better than summarized methods in terms of retaining statistical power, and that the best analytical approach may depend on the pertinent research question and dataset structure. We will expand upon this in the Discussion section. 

\subsection{Examples}
We now present two real-world examples of the utility of the hierarchical bootstrap as applied to behavioral data collected from experiments in songbirds~\cite{hoffmann2014vocal} and flies~\cite{cande2018optogenetic}. These examples provide concrete instances of why one should use the appropriate statistical tests depending on the nature of their data and how the popular tests can result in more false positives or less statistical power than one desires.

\subsubsection{The bootstrap highlights the risk of false positives when analyzing hierarchical behavioral data (vocal generalization in songbirds) using traditional statistical methods}
As described above, although the bootstrap provides a better compromise between statistical power and false-positive rate than the Traditional or Summarized methods, its use is not widespread in the neuroscience literature, including in some of our own prior published work. To illustrate the practical importance of these issues, and to encourage other authors to critically re-evaluate their prior analyses, we here present a case in which we have used the bootstrap and the LMM to reexamine one of our prior results – which used both Traditional and Summarized methods – and found the choices made can significantly affect the outcome of our analyses. As a reminder, when discussing Traditional, Summarized or LMM statistical tests, we will report a p-value denoted by ‘p’ which yields a significant result if p < 0.05. When talking about the Bootstrap tests however, we will report a p\textsubscript{boot} which in turn yields a significant result if p\textsubscript{boot} < 0.025 or p\textsubscript{boot} > 0.975. In addition, p\textsubscript{boot} provides a direct probability of the hypothesis being true, which is not true of p-values~\cite{halsey2015fickle}.

\begin{figure}[!ht]
    \centering
    \includegraphics[width=14cm]{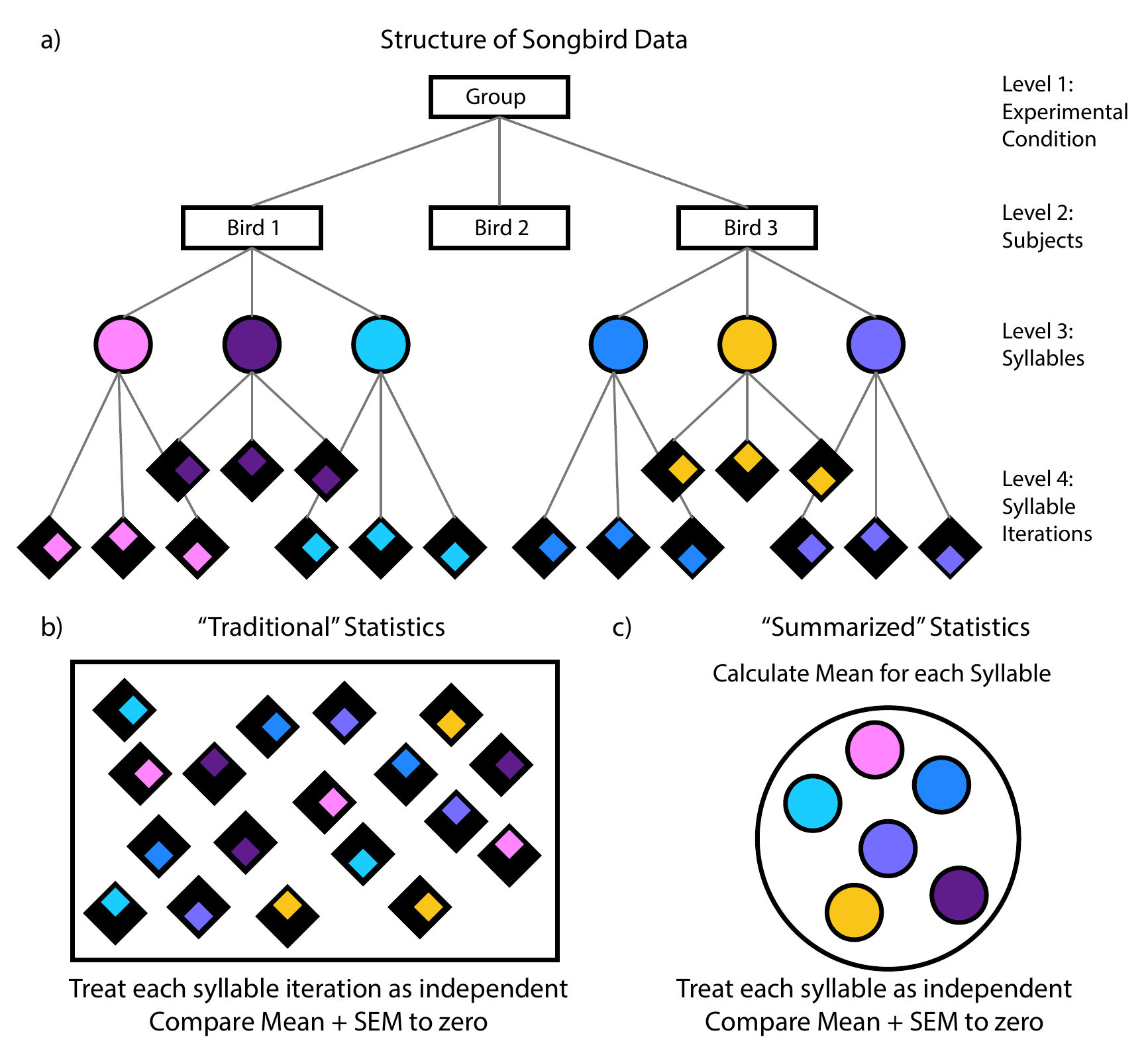}
    \caption{Overview of the structure of songbird data and statistical methods used to analyze it. a) A graphical representation of the hierarchical structure of songbird data. As shown, there are typically multiple experimental groups consisting of several birds each. Each bird in turn sings several syllables and each syllable may be sung several hundreds of times (called iterations). b) In “Traditional” statistics, each syllable iteration is treated as independent and therefore all syllable iterations across syllables and birds are pooled in order to get the population statistics which is then used to perform statistical tests. c) In “Summarized” statistics, the mean for each syllable is calculated from the syllable iterations and the resulting means are pooled across birds in the experimental group. This population of means is then used to perform statistical tests. As detailed in the results, neither of these approaches is valid since even in the “Summarized” case, we are not ensuring independence between data points since syllables within birds are still more similar than syllables between birds.}
    \label{fig:BirdsDataStr}
\end{figure}

\begin{figure}
    \centering
    \includegraphics[width=8cm]{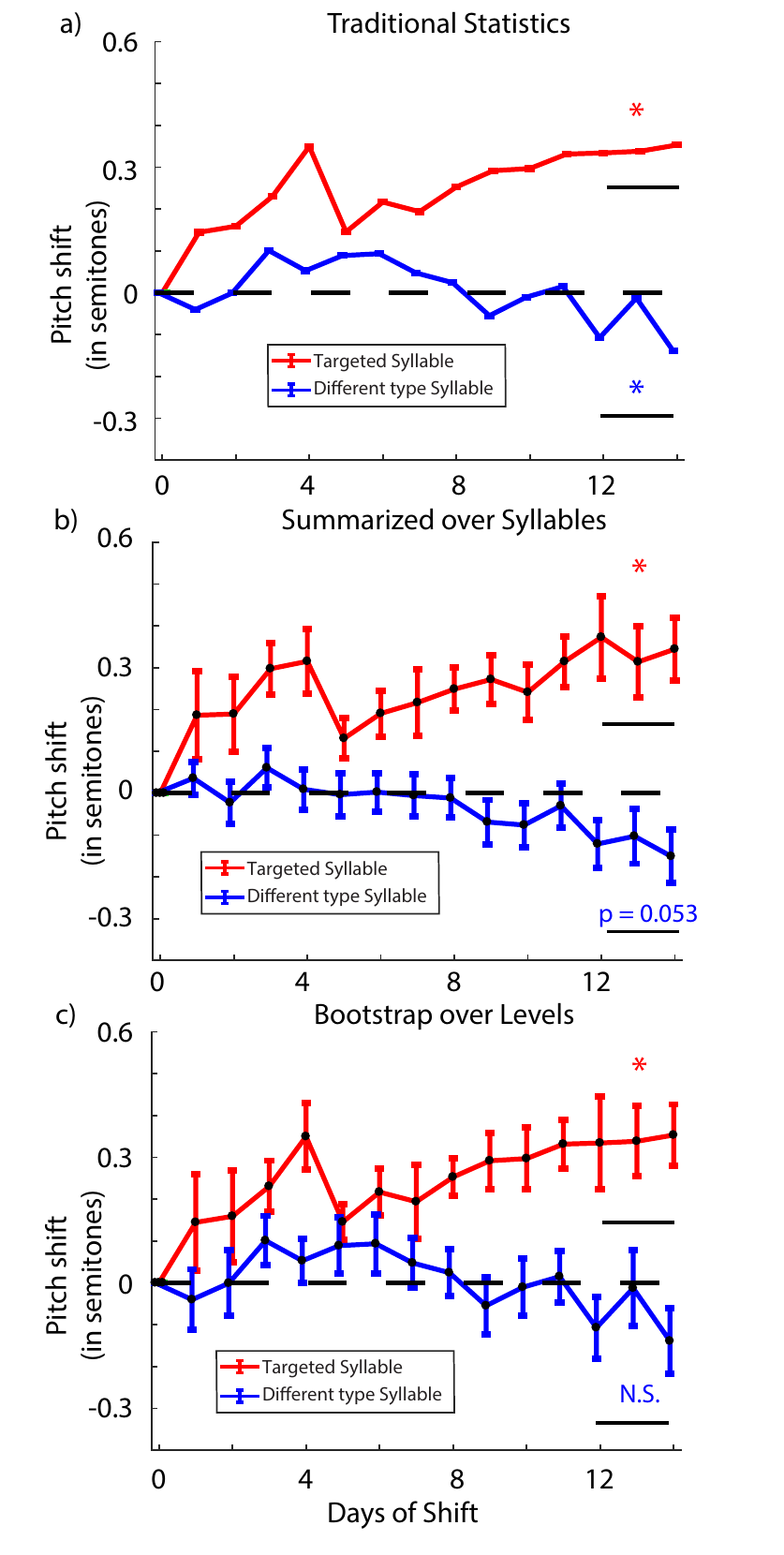}
    \caption{Reanalysis of generalization in the headphones learning paradigm. a) The results when quantified using traditional statistics. As shown, both target and different type syllables differ significantly from zero over the last 3 days of the shift with target syllable moving adaptively and the different type syllables showing anti-adaptive generalization respectively. b) The results when quantified using summarized statistics when summarized over the syllables. In this case, the target syllable is significantly different from zero while the different type syllables are just over the threshold for significance. c) The results when bootstrap is applied over the hierarchical levels. The target syllable is significantly different from zero but the different type syllables are not.}
    \label{fig:BirdsReanal}
\end{figure}

For our first example, in songbirds, each bird sings a variety of syllables and each syllable is repeated a different number of times. A graphical representation of this data structure and the “Traditional” and “Summarized” methods we deployed to analyze this data is shown in Figure~\ref{fig:BirdsDataStr}. As depicted in Figure~\ref{fig:BirdsDataStr}c, we decided to summarize at the level of syllables even though one must summarize at the level of birds in order to ensure independence between data points. In many of our studies, we examine changes in the pitch of these syllables in response to manipulations. In a prior study, we examined the "generalization" (defined below) of vocal learning in songbirds in response to an induced auditory pitch shift on a particular (target) syllable~\cite{hoffmann2014vocal}. In these studies, the auditory feedback of one syllable was shifted in pitch and relayed to the bird through custom-built headphones with very short (~10 ms) latency, effectively replacing the bird’s natural auditory feedback with the manipulated version~\cite{sober2009adult,hoffmann2012lightweight,hoffmann2014vocal}. Note that while the headphones provided auditory feedback throughout the song, only the feedback for a single targeted syllable was shifted in pitch. We reported that in addition to birds changing the pitch of the target syllable in response to the pitch shift, the birds also "generalized" by changing the pitch of other syllables that had not been shifted in pitch. Specifically, we reported that syllables of the same-type (acoustic structure) as the target syllable changed pitch in the same direction as the target syllable (“generalization”) while syllables of a different-type than the target syllable changed pitch in the direction opposite to that of the target syllable (“anti-adaptive generalization”; see Fig.~\ref{fig:BirdsReanal}a). Since in Hoffmann and Sober (2014) we employed traditional and summarized (at a syllable level) statistics when analyzing generalization, we decided to reanalyze the data from that study to ask if the generalization observed was still statistically significant when statistical tests were computed using the hierarchical bootstrapping procedure. In order to do so, we first recapitulated the results reported by computing statistics on the last 3 days of the shift period using the traditional and summarized methods as was reported earlier~\cite{hoffmann2014vocal}. We focus our reporting on changes in the target syllable and anti-adaptive generalization in different-type syllables for the purpose of this example.

When we computed the change in pitch over the last 3 days of the shift period for the syllable targeted with auditory pitch shifts, we found that the birds compensated for the pitch shift of 1 semitone by 0.341 $\pm $ 0.007 (mean $\pm $ SEM in all cases) semitones with traditional statistics (one sampled t-test comparing to zero; t = 47.3; p < 2*10\textsuperscript{-308}; limit due to smallest number in MATLAB; red trace in Fig.~\ref{fig:BirdsReanal}a) and by 0.34 $\pm $ 0.08 semitones with summarized statistics (one sampled t-test comparing to zero; t = 4.25; p = 0.004; red trace in Fig.~\ref{fig:BirdsReanal}b). We did see anti-adaptive generalization in different-type syllables of -0.087 $\pm $ 0.003 semitones with traditional statistics (one sampled t-test; t = 23.9; p = 4*10\textsuperscript{-125}; blue trace in Fig.~\ref{fig:BirdsReanal}a). With summarized statistics, the different-type syllables changed by -0.12 $\pm $ 0.06 semitones (one sampled t-test; t = 2.00; p = 0.053; blue trace in Fig.~\ref{fig:BirdsReanal}b) and was (just) not statistically significant. We note a minor confound in our reanalysis: due to a discrepancy in our data archiving, our recapitulation of the old analysis for the paper yielded a slightly different p-value (p = 0.053) for summarized analysis of different type syllables than was originally reported in the original paper (p =0.048). The point of this analysis is therefore not to replicate the exact findings but to highlight how choices made for statistical analyses can define the interpretation of one’s results as we detail below.

When we reanalyzed the data using bootstrapping over the hierarchical levels, we found that we did not have enough statistical power to claim that the anti-adaptive generalization was statistically significant. As expected, the targeted syllable shifted significantly away from zero to a final value of 0.34 $\pm $ 0.12 semitones (probability of resampled mean being greater than or equal to zero was p\textsubscript{boot} = 0.995; red trace in Fig.~\ref{fig:BirdsReanal}c). As a reminder, p\textsubscript{boot} gives the probability of the hypothesis tested being true. Therefore, a value of 0.5 indicates minimal support for either the hypothesis (or its opposite) while values close to 1 (or 0) represent strong support for (or for the opposite of) the hypothesis. Different-type syllables however shifted to a final value of -0.09 $\pm $ 0.09 (probability of resampled mean being greater than or equal to zero was p\textsubscript{boot} = 0.25; blue trace in Fig.~\ref{fig:BirdsReanal}c). Hence, this result shows that the anti-adaptive generalization was too small an effect to detect with the sample size in the original study, suggesting that the generalization effects observed were driven largely by a small number of individual birds rather than a population wide effect. This conclusion is consistent with an independent study in songbirds that also did not find evidence of anti-adaptive generalization in different type syllables~\cite{tian2017discrete}.

We further reanalyzed the data using LMMs. In order to do so, since we were not interested in the effect of the type of syllable but whether each type of syllable was significantly different from zero, we built three separate LMMs (one for each type of syllable). The model had no fixed effects and the random effects controlled for bird identity and syllable identity within birds. We assessed the intercept and whether it was significantly different from zero. As expected, our results indicated that the target syllable shifted robustly by 0.33 $\pm $ 0.08 semitones (t = 4.24; p = 2.3 * 10\textsuperscript{-5}). Curiously however, the LMMs suggested that different type syllables also shifted robustly by -0.12 $\pm $ 0.06 semitones (t = 2.10; p = 0.036). 

Though omitted here for brevity, we also reanalyzed generalization in same-type syllables and similarly did not find a significant effect using either bootstrapping (probability of resampled mean being greater than or equal to zero was p\textsubscript{boot} = 0.85) or LMMs (t = 1.31; p = 0.19). These results again indicate that we did not perform the generalization experiment on a sufficient number of birds to adequately power the study. However it is worth noting that while the results did not meet the threshold for statistical significance, reporting probabilities in support of the hypotheses (p\textsubscript{boot}) provides more information than simply determining whether or not a statistical threshold was met. In this case, a p\textsubscript{boot} of 0.85 means that if we measured data from more birds drawn from the same distribution, we will see adaptive generalization in 85\% of cases which is much higher than chance (50\%) and is still useful information. Furthermore, we will note that an independent study~\cite{tian2017discrete} did find evidence of adaptive generalization for same type syllables in songbirds.

\subsubsection{The hierarchical bootstrap captures the signal better than traditional or summarized methods in optogenetic control of behavior in flies}
\begin{figure}
    \centering
    \includegraphics[width=9cm]{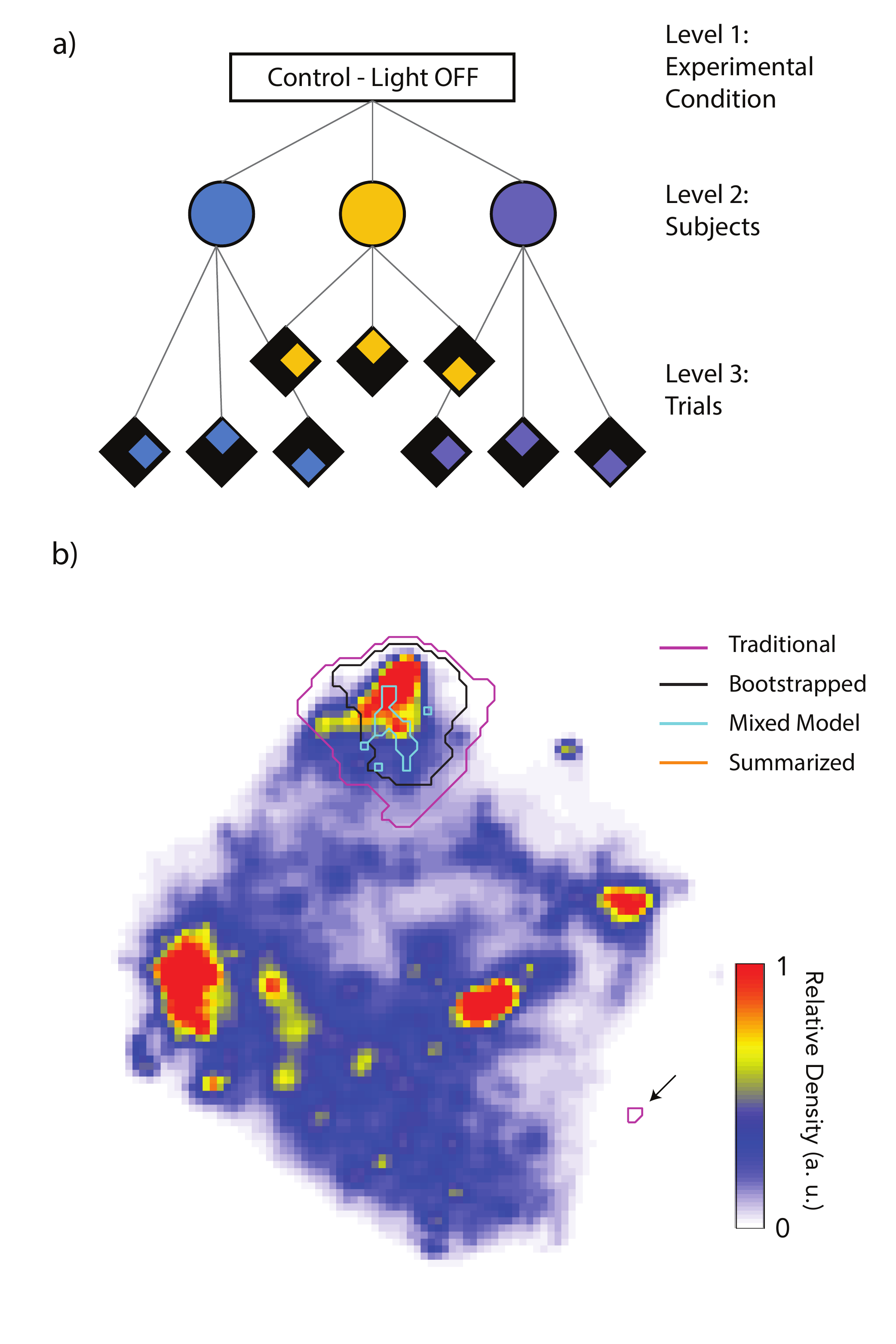}
    \caption{a) A graphical representation of the hierarchical structure present in each experimental group in the study. As shown, each experimental group consists of several subjects (flies) and each fly is observed performing several hundreds to thousands of trials. b) The frequency of behavior mapped onto a two-dimensional space when a particular group of descending neurons in the experimental flies were manipulated using optogenetic stimulation. In this particular case, the descending neurons targeted controlled head grooming behavior, represented at the top of the map. Thus, the animals display elevated frequencies of head grooming during stimulation when compared to control flies and when the light was turned off. The magenta trace shows the statistically significant differences after accounting for multiple comparisons when using the traditional method. As shown, the magenta trace overestimates the signal present and captures some false regions as well as pointed to in the lower right region. The black trace represents the areas of significant difference as defined by using the hierarchical bootstrap and it matches the region we expected in our original analysis very well. The summarized method did not return any regions that were statistically different between groups even though there was a clear signal present in the data (see videos and other data in~\cite{cande2018optogenetic}). The LMM returned an area (cyan trace) that was more conservative than any other method and likely included many false negatives, though the false positive rate was small.}
    \label{fig:FlyOpto}
\end{figure}

We wanted to test the utility of the hierarchical bootstrap in an example where the recorded variables are multi-dimensional, and so we chose to analyze the data from an experiment studying the role of descending neurons in the control of behavior in flies~\cite{cande2018optogenetic}. Studies examining optogenetic effects on behavior are another area where hierarchical datasets are the norm. In this example, since each fly can be tracked over extended periods of time, it will exhibit each behavior multiple times within the period of observation. Additionally, multiple flies can be tracked simultaneously and the behavior is typically averaged across flies across trials for each experimental group. In this study, we used optogenetics to activate descending neurons in flies and studied the corresponding changes in behavior displayed. In order to do so, we first created a two-dimensional representation of the behavior of the flies, as described in detail previously~\cite{berman2014mapping,cande2018optogenetic}. This representation is a probability density function, with each individual peak corresponding to a distinct stereotyped behavior (e.g., running at a given speed, left wing grooming, antennal grooming, etc.). We then mapped the behavior of both experimental animals and control animals (where there was not a functional light-activated channel present) in the presence and absence of light stimulation, onto the behavioral representation. The resulting behavioral map for one class of descending neurons (neurons G7 and G8, line SS02635, from Figure 2 and Videos 6 and 7 in the original study) is shown in Figure~\ref{fig:FlyOpto}b. Through both statistical analysis and visual observation, we found that this stimulating this neuron robustly demonstrated a head grooming phenotype (the upper-most section of the map). 

In the original study, to assess whether the light stimulation caused a statistically significant change in the frequency of behavior observed, we argued that for a region to be significantly increased by the activation, (i) the frequency of behavior had to be significantly greater during optical stimulation than during periods of no stimulation within experimental animals and (ii) the frequency of behavior during optical stimulation had to be greater in experimental animals than in control animals. We used Wilcoxon rank-sum tests coupled with Sidak corrections~\cite{vsidak1967rectangular} for multiple comparisons across the behavioral space in order to test for statistically significant differences for both (i) and (ii).  Full details of this approach are enumerated in the original study~\cite{cande2018optogenetic}, but the number of comparisons is conservatively estimated to be 2\textsuperscript{(H(ρ))}, where H(ρ) is the entropy of the two-dimensional probability distribution. This approach would fall under the category of traditional methods that we have described previously. We compared the regions obtained from the original analysis with regions we obtained when using summarized or bootstrap statistics on this dataset and the result is shown in Figure~\ref{fig:FlyOpto}b. As shown, the traditional method seems to overestimate the region of significant differences and includes a false positive area that is separate from the main region where signal is present. We are confident that this area is a false positive because all trials were individually visually inspected, with each one displaying the head grooming phenotype when stimulated. 

The summarized method, on the other hand, does not identify any regions as being statistically significant, despite visual evidence suggesting the clear presence of a signal~\cite{cande2018optogenetic}. This lack of an identified region likely results from the fact that the number of individual flies used was not high (six experimental flies and six control flies), but with 30 trials for each animal. Thus, the summarized method did not sufficiently account for the robust trial-to-trial response of each animal. As is the case for any hierarchical data set, the effective number of samples is not six, but nor is it 180 (six animals multiplied by 30 trials) because of the correlations between subsequent trials in the same animal. Taking these correlations into account, however, the hierarchical bootstrap returns a concise region that is more tightly fit around the region that is associated with head grooming in non-stimulated animals (Video 2 in the original study). Thus, even in this case of abstracted, multi-dimensional data analysis, we see that the hierarchical bootstrap is an effective method that accounts for correlation structure within a dataset.  

We also performed a LMM to test in a similar manner to the birdsong example described previously.  Specifically, we assumed no fixed effects, controlled random effects for fly identity, and we tested for the significance of the intercept (using the same multiple hypothsis testing threshold as described above). We performed this analysis in a one-sided manner for both the control and the experimental animals, identifying regions as “significant” if the fitted intercept for the experimental animals was positive with a sufficiently small p-value and if the control animals did not experience a significant positive shift. As seen in the cyan line of Figure~\ref{fig:FlyOpto}b, we see that the LMM was more conservative than the Bootstrap method in this instance, picking a smaller, somewhat discontinuous, portion of the map. Based on observations from Video 2 of the original study, the LMM likely is resulting in many false negatives, however the false positive rate is low. 

\section{Discussion}
The hierarchical bootstrap is a powerful statistical tool that was developed to quantify the relationship between school class sizes and achievement~\cite{carpenter2003novel} and has since been used to quantify effect sizes in a wide variety of fields. The use of the hierarchical bootstrap in neuroscience however is still limited in spite of the need for it being increasingly clear. Through our simulations, we have shown the utility of the hierarchical bootstrap by examining the shortcomings of more common statistical techniques in a typical example one might encounter. While the results of our simulations may be inferred from other mathematical work on the subject~\cite{davison1997bootstrap,carpenter2003novel,field2007bootstrapping,goldstein2011bootstrapping}, our goal here is to illustrate the application of these methods to the hierarchical datasets typically found in neuroscience. We first illustrated that the bootstrap does not have a conservative bias for hierarchical datasets in which the assumption of independence between data points is violated (Fig.~\ref{fig:NullDep}). We then showed that the bootstrap performs better than summarized statistical measures by not sacrificing as much statistical power especially at low sample sizes and small effect sizes (Fig.~\ref{fig:Power}). Finally, we showed real world applications of applying the hierarchical bootstrap to two datasets from songbirds and flies to demonstrate the advantages of the hierarchical bootstrap over other more commonly used statistical analysis techniques in neuroscience (Figs.~\ref{fig:BirdsReanal} and~\ref{fig:FlyOpto}).

A critical assumption of several commonly used statistical tests is that every data point within a dataset is independent. In other words, every data point would have to be equally different from every other data point for any two randomly selected data points. “Pseudoreplication” refers to studies in which individual data points may be more similar to some data points than to others within the dataset, and yet the statistical tests performed treated the data points as independent replicates~\cite{hurlbert1984pseudoreplication}. While pseudoreplication was first extensively reported on in ecological studies~\cite{hurlbert1984pseudoreplication,heffner1996pseudoreplication}, it has since been identified as a common problem in other fields, including neuroscience~\cite{lazic2010problem}. While resampling methods including bootstrapping were originally suggested as tools by which one could overcome the pseudoreplication problem~\cite{crowley1992resampling}, the bootstrap was argued to have a conservative bias resulting in larger error bars than necessary~\cite{hillis1993empirical,adams1997resampling}. Since then, however, several versions of the bootstrap algorithm have been developed to apply to hierarchical data and have been found to be unbiased and more robust for calculation of uncertainty in clustered data than other statistical methods~\cite{davison1997bootstrap,carpenter2003novel,field2007bootstrapping,goldstein2011bootstrapping,harden2011bootstrap,huang2018using}. In order to test the bootstrap for any potential bias in a typical example we might encounter in neuroscience, we performed simulations to quantify differences in mean firing rates between two groups of neurons when there was no difference between the groups. We illustrated that the bootstrap produced a false positive rate significantly below the expected 5\% (Fig.~\ref{fig:NullIndep}b) and had larger error bars (Fig.~\ref{fig:NullIndep}c) than other statistical methods when the data were independent. However, when the independence between data points was abolished by introducing a hierarchical structure, the bootstrap was not statistically different from the expected 5\% false positive rate (green bars in Fig.~\ref{fig:NullDep}b) and the error bars computed were similar to those computed using summarized statistics (red and green bars in Fig.~\ref{fig:NullDep}c) demonstrating that the hierarchical bootstrap is robust to bias for applications in neuroscience.

Furthermore, we showed that the bootstrap does not lose power to the degree that using summarized statistics does (see green traces versus red traces in Fig.~\ref{fig:Power}c) while also keeping the false positive rate within the intended bound (see green trace in Fig.~\ref{fig:Power}b). Additionally, the hierarchical bootstrap as applied in this paper does not make assumptions about the relationships underlying the latent variables that define the hierarchical structure. However, as we saw in Figure~\ref{fig:Power}b, where we sampled multiple trials from a very small number of neurons, the bootstrap assumes that the data collected captures essential characteristics of the population distribution. In this case, the bootstrap initially considered the trial level data as independent before switching to neuron level data as the number of neurons increased. Hence, one may have to adjust the resampling procedure to ensure that the distribution of resampled data points most accurately matches the population distribution one wishes to study.

Among the reasons Linear Mixed Models (LMMs) gained in popularity for statistical testing was the fact that they could accommodate hierarchical datasets by controlling for various levels as “random” effects while still using all available data, thereby minimizing the loss of statistical power~\cite{snijders1993standard,diez2002glossary,snijders2011multilevel,aarts2014solution,hox2017multilevel}. As shown in Figures~\ref{fig:NullDep} and ~\ref{fig:NullIndep}, LMMs do not exhibit a conservative bias for independent or hierarchical datasets. In the particular example we chose for our simulations, LMMs seem to perform better than the bootstrap in terms of eliminating false positives but underperform slightly in retaining statistical power (see magenta versus green traces in Fig.~\ref{fig:Power}b and c). Given that LMMs are considered by many to be the gold-standard for hierarchical datasets~\cite{baayen2008mixed,aarts2014solution,koerner2017application}, why are we advocating an alternative approach?

Given the results from the simulations, we see the bootstrap as a method that performs with a similar rate of false-postitives and false-negatives as the LMM, but requires fewer implementation choices by the user. Unlike the bootstrap, LMMs require the assumptions of the structure in the dataset to be specified in the terms used to model the “random effects”. If this procedure is performed incorrectly, the results obtained and inferences drawn may be inaccurate~\cite{barr2013random,matuschek2017balancing,seedorff2019maybe}. Furthermore, there are some types of hierarchical structures, such as when the relationship between levels is non-linear, that LMMs may not be able to capture the dynamics of the data accurately~\cite{eager2017mixed}. Given these complications and the steep learning curve associated with using LMMs accurately, it has been argued that a perfectly implemented ANOVA may be more useful than a hastily implemented LMM~\cite{seedorff2019maybe}. A final danger associated with learning to use LMMs correctly that it could encourage p-hacking since one could potentially obtain the answer they desire by changing the structure of the random effects.

To showcase its utility in analyzing hierarchical datasets in neuroscience, we then used the hierarchical bootstrap on two independent examples. First, we reanalyzed data from Hoffmann and Sober, 2014 in which we used both Traditional and Summarized statistical analysis to conclude that songbirds generalize changes in pitch targeted on a single syllable anti-adaptively to syllables of a different type. When reanalyzed with the bootstrap however, we found that the anti-adaptive generalization of different type syllables (blue trace in Fig.~\ref{fig:BirdsReanal}c) did not meet the threshold for statistical significance. This was a striking result, as the original study did report statistically significant changes from zero even while using summarized statistics~\cite{hoffmann2014vocal}. A probable reason for the differences between the summarized and bootstrap methods for this dataset stems from a decision point regarding the level to which one must summarize the data when using summarized statistics (we avoided this decision in the simulations by assuming all data came from a single subject). The summary statistics reported were summarized at the level of syllables for this dataset. However, in order to truly make sure all points are independent, one must summarize at the highest level, i.e., at the level of individual birds in the dataset. The differences in results between the summarized and bootstrap methods here suggest that the generalization effects were driven largely by strong effects in a subset of birds as opposed to a population-wide effect and that, by failing to take the hierarchical nature of the dataset into account, we overestimated our statistical power and chose too low an N. Further evidence for this interpretation, as opposed to one where incorrect analysis methods call all former results into question, comes from reanalysis of data from a separate study looking at learning birds display in response to pitch shift of their entire song through custom-built headphones~\cite{sober2009adult} using the hierarchical bootstrap. Since the changes in pitch were far more systemic across birds in this experiment, we did not see any changes in statistically significant results~\cite{saravanan2019dopamine}. 

This example also served to highlight the complexities associated with implementing and interpreting results from an LMM accurately. When we reanalyzed the songbird data using LMMs we were surprised to see a significant effect of anti-adaptive generalization in different type syllables but not for adaptive generalization in same type syllables. While it is possible that this result may be true in our particular dataset it is highly unlikely as the results are the opposite of what other studies have reported in songbirds~\cite{tian2017discrete} and humans~\cite{pile2007talking,villacorta2007sensorimotor}. This suggests that in spite of our best efforts to accurately specify LMMs to fit this data, we may have failed in capturing all the essential structure required in the random effects.

In addition to re-examining published data from vocal behavior in songbirds, we applied the hierarchical bootstrap to data from an independent experiment studying the role of descending neurons in controlling behavior in flies using optogenetics~\cite{cande2018optogenetic}. As shown in Figure~\ref{fig:FlyOpto}b, the hierarchical bootstrap performs better than the traditional and summarized statistical methods and LMMs in isolating the true signal in the experiment. The traditional method includes areas that are likely false positives and the summarized method does not return any statistically significant areas. The LMM was interestingly much more conservative than the bootstrap in this case and probably had a much larger false negative rate though the false positive rate was low.

We would also like to reiterate another advantage of the direct probabilities returned by the bootstrap (p\textsubscript{boot}) over p-values typically reported through the traditional, summarized, and LMM approaches. p-values represent the probability of obtaining results at least as extreme as the ones obtained under the assumption that the null hypothesis is true. It is a cumbersome definition that has led to numerous misconceptions regarding what p-values actually signify~\cite{halsey2015fickle,wasserstein2016asa}. The value returned by the bootstrap however, p\textsubscript{boot}, provides a direct probability in support of a particular hypothesis. As we reported in the songbirds example, we found that the probability of same-type syllables generalizing was 0.85. This means that if we measured data from more birds drawn from the same distribution, we will see adaptive generalization in 85\% of cases which is much higher than chance (50\%). Hence, the hierarchical bootstrap method can provide a measure of the relative support for the hypothesis which is both easier to understand and can be useful information for both positive and negative results in research.

To conclude, neuroscience research is at an exciting junction. On the one hand, new technologies are being built promising bigger and more complex datasets to help understand brain function~\cite{yizhar2011optogenetics,burns2013open,vogelstein2018community}. On the other, we have rising concerns over the incorrect use of statistical tests~\cite{ioannidis2005most,nieuwenhuis2011erroneous,greenland2016statistical} and the lack of reproducibility of a number of past findings~\cite{baker20161,milkowski2018replicability,gerlai2019reproducibility}. We propose the hierarchical bootstrap as a powerful but easy-to-implement method that can be scaled to large and complicated datasets, that returns a direct probability in support of a tested hypothesis reducing the potential for misinterpretation of p-values and that can be checked for correct implementation through sharing of analysis code. As we have shown through this paper, we believe that widespread use of the bootstrap will reduce the rate of false positive results and improve the use of appropriate statistical tests across many types of neuroscience data.

\section*{Acknowledgements}
The work for this project was funded by NIH NINDS F31 NS100406, NIH NINDS R01 NS084844, NIH NIBIB R01 EB022872, NIH NIMH R01 MH115831-01, NSF 1456912, Research Corporation for Science Advancement no. 25999 and The Simons Foundation.

\section*{Conflict of Interest}
The authors declare no competing financial interests.

\printendnotes

\bibliography{sample}

\end{document}